\documentclass{aa}  

\usepackage{graphicx}
\usepackage{txfonts}
\begin{document}

   \title{A living catalog of stream interaction regions in the Parker Solar Probe era}

   \author{R. C. Allen\inst{1}
          \and
          G. C. Ho\inst{1}
          \and
          L. K. Jian\inst{2}
          \and
          S. K. Vines\inst{1}
          \and
          S. D. Bale\inst{3,4,5,6}
          \and
          A. W. Case\inst{7}
          \and
          M. E. Hill\inst{1}
          \and
          C. J. Joyce\inst{8}
          \and
          J. C. Kasper\inst{7,9}
          \and
          K. E. Korreck\inst{7}
          \and
          D. M. Malaspina\inst{10}
          \and
          D. J. McComas\inst{8}
          \and
          R. McNutt\inst{1}
          \and
          C. M\"{o}stl\inst{11}
          \and
          D. Odstrcil\inst{2, 12}
          \and
          N. Raouafi\inst{1}
          \and
          N. A. Schwadron\inst{13}
          \and
          M. L. Stevens\inst{7}
          }

   \institute{Johns Hopkins Applied Physics Lab, Laurel, MD 20723, USA.\\
        \email{Robert.Allen@jhuapl.edu}
         \and 
         NASA Goddard Space Flight Center, Greenbelt, MD 20771, USA.
         \and
         Physics Department, University of California, Berkeley, CA 94720, USA
         \and
         Space Sciences Laboratory, University of California, Berkeley, CA 94720, USA
         \and
         The Blackett Laboratory, Imperial College London, London, SW7 2AZ, UK
         \and
         School of Physics and Astronomy, Queen Mary University of London, London E1 4NS, UK
         \and
         Smithsonian Astrophysical Observatory, Cambridge, MA 02138, USA
         \and
         Department of Astrophysical Science, Princeton University, Princeton, NJ, USA
         \and
         Climate and Space Sciences and Engineering, University of Michigan, Ann Arbor, MI 48109, USA
         \and
         Laboratory for Atmospheric and Space Physics, University of Colorado, Boulder, CO 80303, USA
         \and
         Space Research Institute, Austrian Academy of Sciences, Graz, Austria
         \and
         Physics and Astronomy Department, George Mason University, Fairfax, VA 22030, USA
         \and
         University of New Hampshire, Space Science Center, Durham, NH, USA}

   \date{Received 3 November 2020; accepted 7 December 2020}

 
  \abstract
   {Stream interaction regions (SIRs) and corotating interaction regions (CIRs) are important phenomena in heliospheric physics. These large-scale structures vary temporally and spatially, both in latitude and with radial distance. The additions of Parker Solar Probe (PSP) and Solar Orbiter have allowed for investigations into the radial evolution of these structures over a wide range of heliocentric distances for the first time since the Helios era. To better enable investigations of SIRs and CIRs within the inner heliosphere, we have developed a living catalog of SIR and CIR observations by Parker Solar Probe with corresponding observations by STEREO-A as well as ACE and Wind at 1 au. The methodology used for the identification of events and the generation of this catalog, the initial catalog of PSP observations spanning orbits one through five along with corresponding 1 au observations, and information on accessing the living catalog for future studies is described. This list of SIR and CIR events from PSP and corresponding observations from other heliophysics missions will enable case studies utilizing unique orbital arrangements, as well as aid in future statistical studies to further understand the properties and evolution of these structures.}

   \keywords{Catalogs --
                Sun: Solar Wind --
                acceleration of partices
               }

   \maketitle
%

\section{Introduction}

Stream interaction regions (SIRs) form in interplanetary space where faster solar wind streams, originating from coronal holes, overtake the preceding slower solar wind streams (e.g., \citealp{Belcher1971, Pizzo1978, Richardson2018}). Where the two streams meet, a density "pileup" of compressed plasma is formed upstream of the stream interface, which is typically identified as the peak in total pressure between the two solar wind populations, and is followed by a rarefaction region in the fast solar wind (see \citealp[and references therein]{Richardson2018}). As the SIR propagates away from the Sun, the compression can lead to the formation of an SIR-associated forward and reverse shock pair at and often beyond 1 au, which efficiently accelerates particles (see \citealp[and references therein]{Jian2006, Jian2008a, Richardson2018}). As such, SIRs have been found to be a major source of energetic particles in interplanetary space (e.g., \citealp{VanHollebeke1981, Tsurutani1982, Richardson2018, Kollmann2019}). As the main source of the high-speed stream, the coronal hole rotates around the solar rotation axis, and the SIR structure rotates with it. After a complete solar rotation, the SIR is classified as a corotating interaction region (CIR). SIRs and CIRs can also be highly geoeffective, triggering geomagnetic storms (e.g., \citealp{Tsurutani1997, Turner2006}) and affecting the ionosphere and thermosphere of Earth (e.g., \citealp{Chen2014}). Additionally, these structures, and their associated shocks, can modulate the intensity of galactic cosmic rays from the inner heliosphere out to the very local interstellar medium (e.g., \citealp{Hill2020a}).

The properties of SIRS and CIRs vary with radial distance and latitude, as well as over the lifetime of the SIR and CIR as the footpoint of the structure rotates around the solar rotation axis. Prior to the launch of Helios 1 and 2 in 1974, \citet{Hundhausen1973} constructed a hydrodynamic model of the formation of SIRs in the inner heliosphere. The model predicted that as the fast streams flow away from the Sun, the velocity transition from slow-to-fast wind would steepen, the density pileup upstream of the SIR would grow, and the pressure peak and temperature increase near the interface would both become relatively stronger and sharper \citep{Hundhausen1973}. These modeling results were later confirmed by superposed epoch investigations of SIRs where events between 0.3-0.4 au at 0.9-1 au were compared \citep{Richter1986}. Later comparisons of SIR observations at Pioneer Venus Orbiter (at 0.7 au) and those at 1 au expanded upon these studies, finding that the rate of SIRs with sharp stream interfaces increased from 9\% at 0.7 au to 21\% at 1 au \citep{Jian2008b}. 

Regarding the intensity of energetic particles associated with SIRs at different radial distances, \citet{VanHollebeke1978} compared SIR and CIR events between Helios 1 and 2 (spanning 0.3-1 au), IMP 7 (at 1 au), Pioneer 10 (spanning 9-10 au), and Pioneer 11 (at 3.8 au). These comparisons revealed that the SIR- and CIR-associated suprathermal flux increased in intensity with radial distance from the Sun to \textasciitilde3 au, where the intensity peaked, before decreasing again at larger radial distances \citep{VanHollebeke1978}. This motivated the current understanding that most SIR- and CIR-associated shocks form around 3 au (see \citealp[and references therein]{Richardson2018}). Particles accelerated at these shock fronts can propagate back into the inner heliosphere along magnetic field lines, although scattering processes and magnetic cooling decrease the intensity and change the spectral form of the particles as they approach the Sun (e.g., \citealp{Mewaldt1978, Fisk1980, Mason1999, Zhao2016}). In addition to radial variations in the ion spectra, the suprathermal ion spectra and intensity can also be different between spacecraft closely separated in longitude, depending on where along the shock surface at greater heliospheric distances the particles are magnetically connected (e.g., \citealp{Zhao2015}).

Along with particles accelerated further out in the heliosphere, local, non-shock-related acceleration has been proposed to occur in the inner heliosphere associated with the stream interface (e.g., \citealp{Schwadron1996, Giacalone2002, Ebert2012, Filwett2017, Filwett2019, Allen2020a}). These studies have suggested that in the inner heliosphere, the lower-energy component of SIR- and CIR-associated suprathermal ions may be a result of local acceleration, while the higher-energy component is a result of distant shock acceleration (e.g., \citealp{Ebert2012, Filwett2019}). This has been supported by observations of acceleration at SIR compressions within 1 au, where shocks have not yet formed \citep{Chotoo2000, Chen2015}. Recent observations by Parker Solar Probe (PSP) at 0.3 au \citep{McComas2019} have further suggested the importance of compression-related acceleration at lower heliospheric distances than where SIR- and CIR-associated shocks typically form \citep{Joyce2020b}. PSP observations have also observed weaker modulation of the energetic particles from transport than expected, suggesting sub-Parker field lines may allow a more direct access to shock regions \citep{Joyce2020b, Schwadron2020}. 

In addition to the radial evolution of SIRs and CIRs as well as their associated suprathermal particles, these structures also vary temporally as they corotate with the Sun. The bulk plasma properties and suprathermal ion intensity can vary from rotation to rotation (e.g., \citealp{Mason2009, Allen2020b}) as the initial stream forms into a well-developed CIR with associated shocks and continues to evolve. More generally, SIR and CIR properties as well as occurrences have a solar cycle dependence \citep{Jian2011, Jian2019}, as does the SIR and CIR suprathermal ion composition (e.g., \citealp{Mason2008, Mason2012, Allen2019}). The combination of temporal and radial evolution along with longitudinal, latitudinal, and solar cycle variations underscore the importance of multi-point and successive observations between various missions to better study the dynamics and evolution of SIRs and CIRs. 

With the addition of PSP to the Heliophysics System Observatory (HSO), we have the ability to investigate SIRs and CIRs within the inner heliosphere for the first time since the Helios era (see \citealp{McComas2019, Allen2020a, Allen2020b, Cohen2020, Desai2020, Joyce2020a, Joyce2020b}). To better understand the variability and evolution of SIRs and CIRs, we present here an initial catalog of SIRs and CIRs observed by PSP during the first five orbits of the mission, along with corresponding observations at the Solar Terrestrial Relations Observatory-Ahead (STEREO-A), Advanced Composition Explorer (ACE), and Wind spacecraft. Future iterations of this catalog will also include corresponding SIR and CIR observations from Solar Orbiter \citep{Muller2020} as the data become available. Updates to this catalog will correspond to each new public release of PSP data and will be publicly available through the PSP Science Gateway (https://sppgway.jhuapl.edu/event\_list/). This will allow for researchers to not only find fortuitous events, but also facilitate new statistical studies needed to further our understanding of the spatiotemporal evolution of SIRs and CIRs as well as their associated energetic particle populations. Section 2 discusses the different missions and datasets currently used for this catalog and describes the criteria used to identify SIRs and CIRs for inclusion in this list. Section 3 illustrates the PSP observed SIRs and CIRs over the first five orbits along with corresponding observations by STEREO-A and L1 missions (ACE and Wind). A detailed look at two of these identified SIR and CIR events and preliminary statistical analysis of the radial evolution of SIR and CIR properties are also discussed. Finally, a summary of this catalog, along with a discussion of how it could be used for future investigations, is given in Section 4. 


\section{Methodology}
\subsection{Parker Solar Probe}

The PSP mission \citep{Fox2016} launched into orbit around the Sun on 2018 August 12 and is providing key measurements of the solar wind and interplanetary magnetic field in the inner heliosphere. Over the lifetime of the mission, the perihelion of PSP has gradually lowered through the use of Venus gravity assists. The observations shown here are 1-minute moment averages of bulk solar wind measurements from the Solar Probe Cup (SPC \citealp{Case2020}), which is part of the Solar Wind Electrons Alpha and Protons (SWEAP) instrument suite \citep{Kasper2016}, along with 1-min averages of magnetic field measurements from the FIELDS suite \citep{Bale2016}. To identify SIRs and CIRs, the solar wind tangential flow deflection is computed using the equation $\alpha_{tan} = tan^{-1} (v_T / v_R)$, where $v_T$ and $v_R$ are the tangential and radial velocity components, respectively, in the radial-tangential-normal (RTN) coordinate system. The SPC proton thermal velocity data product is used to obtain a scalar proton temperature through

\begin{equation}
T = \frac{m}{2k_{B}}v_{therm}^{2}
\end{equation}

where $T$ is the temperature, $m$ is the mass of a proton, $k_B$ is Boltzmann's constant, and $v_{therm}$ is the thermal velocity. The specific entropy argument ($S$) of the solar wind is computed from the SPC proton density ($n$) and the derived temperature using the formulation (e.g., \citealp{Jian2006}),

\begin{equation}
S = \ln{\left(\frac{T^{1.5}}{n} \right)}
.\end{equation}

Additionally, this study uses a combined proton plasma and magnetic pressure ($P$) derived from the FIELDS magnetic field magnitude ($B$) and the SPC proton density and derived temperature,

\begin{equation}
P = nkT + \frac{B}{2\mu}
.\end{equation}

To investigate the suprathermal ion abundance during the SIR and CIR events, 30-min averaged measurements from the Energetic Particle Instrument-Lo (EPI-Lo, \citealp{Hill2017, Hill2020b}), which is part of the Integrated Science Investigation of the Sun (IS$\odot$IS, \citealp{McComas2016}), are used. This paper presents only one EPI-Lo data product, the "IonToF" product, which is a time-of-flight only measurement that allows for observations of ions from \textasciitilde30-586 keV nuc\textsuperscript{-1}, but does not distinguish between separate ion species.

\subsection{STEREO-A}

Since the aim of this catalog is to highlight corresponding observations between PSP and other spacecraft in the HSO, observations from the STEREO mission \citep{Kaiser2008}, a set of two spacecraft that were launched on 2006 October 25, are used. However, only observations from STEREO-A (STA) are available in the PSP era. Similar to PSP, 1-minute averaged magnetic field measurements from the magnetometer \citep{Acuna2008} and suprathermal particles from the Solar Electron and Proton Telescope (SEPT, \citealp{Muller2008}) instruments on the In situ Measurements of Particles and CME Transients (IMPACT) investigation \citep{Luhmann2008} are used to identify SIRs and CIRs. SEPT measures ions in the energy range of 60-7000 keV, but it cannot differentiate the ion species. Additionally, 1-minute bulk solar wind properties (i.e., velocity, temperature, and density) from the PLAsma and SupraThermal Ion Composition investigation (PLASTIC, \citealp{Galvin2008}) are used. Calculations of $T$, $S$, and $P$ following the same procedures as for PSP are also performed.

\subsection{L1 Missions: Wind and ACE}

In addition to STA, this catalog links SIR and CIR observations at PSP to those from the Wind mission \citep{Acuna1995}. Wind, which launched in 1994, has been stationed near the Sun-Earth Lagrange point 1 (L1) since 2004. The SIR and CIR search and catalog entries are based on 1-minute resolution bulk solar wind plasma properties from the Solar Wind Experiment (SWE, \citealp{Ogilvie1995}) and magnetic field measurements from the Magnetic Fields Investigation (MFI, \citealp{Lepping1995}). As is done with STA measurements, $T$, $S$, and $P$ are calculated following the same procedures as for PSP.

The last observatory currently included in the catalog is the ACE mission \citep{Stone1998}. ACE, which launched in 1997, is stationed near L1 and is used in tandem with Wind since their observations are nearly identical. The combination of Wind and ACE allows for a more continuous dataset. Solar wind bulk properties (velocity, density, and temperature) are provided by the Solar Wind Electron Proton Alpha Monitor (SWEPAM, \citealp{McComas1998}). ACE magnetic field measurements are provided by the Magnetic Field Experiment (MAG, \citealp{Smith1998}), and suprathermal ion observations are from the Electron, Proton, and Alpha Monitor-Low Energy Magnetic Spectrometer (EPAM-LEMS, \citealp{Gold1998}) instrument. Our analysis uses 64 s averages of SWEPAM observations, 1-min averages of MAG observations, and 5-min ion intensity averages from EPAM/LEMS120.

\subsection{Catalog criteria}

Following the same criteria as \citet{Jian2006, Jian2013}, PSP-observed SIR and CIR events were selected that matched the majority of the following criteria given the available data: (1) velocity increase, (2) total pressure peak observed during the velocity increase, (3) enhancement of density ahead of the interface, (4) local maxima in the magnetic field magnitude during the velocity increase, (5) temperature and entropy increase from the slow to fast solar wind, and (6) a tangential flow deflection around the interface. These events are then categorized into one of the following three subsets identified with quality flags: clean events without data gaps with at least five of the six criteria satisfied (denoted by a value of zero); clean events (i.e., at least five of six criteria satisfied) with data available through the interface region but with gaps in the slow and/or fast solar wind near the velocity increase (value of one); and events missing some desired data with larger data gaps occurring during the velocity increase (value of two), which still satisfy criteria one, three, and five. Examples of these are shown in the following section. It should be noted that these SIR and CIR events are identified with only observations from SWEAP and FIELDS, and as such data are only available for a subset of the orbit. There are intervals, particularly when PSP is further out in its orbit closer to aphelion, where IS$\odot$IS data are available without corresponding observations from FIELDS and SWEAP (see events in \citealp{Cohen2020, Desai2020}). As a result, it is likely that SIR and CIR observations within the IS$\odot$IS dataset are missing from this catalog. Additionally, to focus on clean SIR and CIR observations, events that were determined to occur near interplanetary coronal mass ejections (ICMEs) are not included in this list. As this catalog uses more stringent selection criteria, some of the high-speed streams from \citet{Allen2020a} are not included in this list.

After identifying an SIR and CIR observed by PSP, the structure is tracked using the approach of \citet{Allen2020a}. First, the structure is identified in a global 3D magnetohydrodynamic heliospheric Wang-Sheeley-Arge (WSA)-ENLIL model \citep{Arge2004, Odstrcil2004} using synoptic solar magnetic field maps derived from magnetograms. For the purpose of building this catalog, zero-point-corrected quick-reduce magnetogram synoptic maps from the Global Oscillation Network Group (GONG, \citealp{Harvey1996}) were used. After identifying the potential SIR and CIR structure in the WSA-ENLIL model (found at http://helioweather.net/missions/psp/), the structure was followed both forward and backward in time until it would reach either STA or L1. This coarse timing approximation is then used as a guide in searching the STA or L1 datasets for observations of the structure. If the observational characteristics suggest that an SIR and CIR, which was observed by PSP, was seen by either STA or L1, then the 1 au observation time is included in the list. This process is then repeated until the SIR or CIR is not observed, allowing for the creation of the catalog of SIR and CIR observations that correspond to those seen by PSP. We note that there are other SIR and CIR events that have been listed by others (e.g., \citealp{Jian2019}) which were not observed by PSP, and as such are not included here. Future work will expand on this catalog of SIRs and CIRs as the PSP mission progresses. Additionally, this same process will be expanded to include Solar Orbiter as those data become available. 

\section{Results}

\subsection{PSP SIRs and CIRs}

\begin{figure*}
   \centering
   \includegraphics[width=18.25cm]{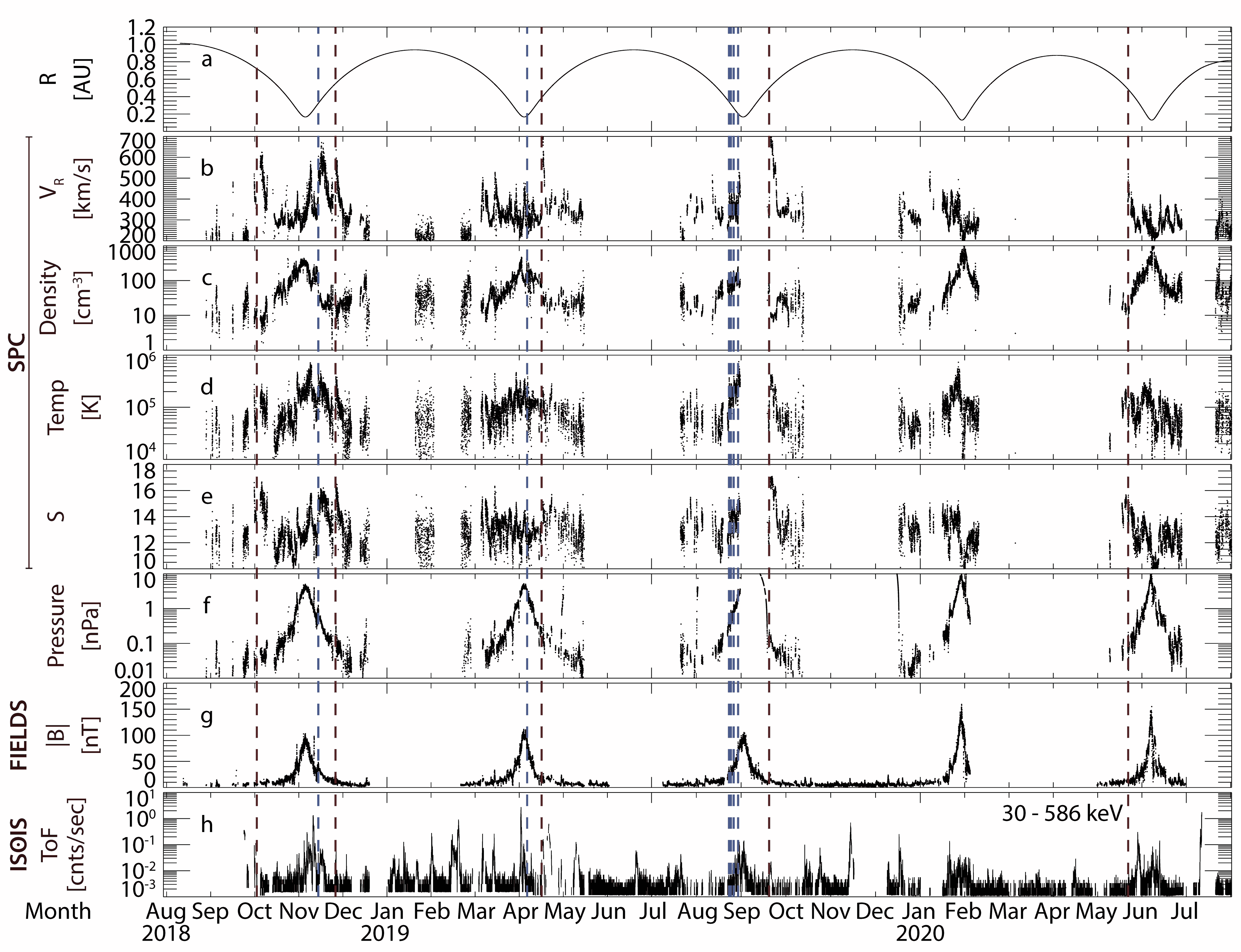}
      \caption{Overview of the first five orbits of PSP showing: (a) heliocentric distance, (b) radial H\textsuperscript{+} velocity, (c) H\textsuperscript{+} density, (d) H\textsuperscript{+} temperature, (e) specific entropy argument, (f) sum of H\textsuperscript{+} plasma and magnetic pressures, (g) magnetic field magnitude, and (h) suprathermal ion count rate. Blue and red dashed lines indicate the times of clean and ambiguous SIR and CIR observations, respectively.
              }
         \label{fig1}
   \end{figure*}

In situ observations from PSP during the first five orbits are shown in Figure \ref{fig1}. During these orbits, 11 SIRs and CIRs were observed that met our above criteria, and they are denoted by color coded vertical lines (Figure \ref{fig1}). Data gaps often occur when PSP is at larger heliospheric distances, especially for SPC, as well as during Ka-band downlinking intervals, and so this limits the time periods that are searchable for SIRs and CIRs. Following the methodology listed in Section 2.4, no SIRs or CIRs were observed during orbit four. This is likely due to PSP traveling fast enough around the Sun so that it likely remained between high-speed streams emanating from coronal holes during this time period (not shown). Apparent velocity enhancements that are not flagged as SIR and CIR events in Figure \ref{fig1} are found to correspond to ICMEs, sector boundary crossings, or slow velocity increases without the formation of a stream interface. 

Table \ref{tab1} lists the eleven SIRs and CIRs observed at PSP during the first five orbits. The event numbers listed in the first column are used for later reference with the identified corresponding events. The SIR and CIR start and end times are identified by eye based on the plasma and fields measurements. Events that occur near data gaps may not have fully accurate start and/or stop times identified. The interface time is defined as the time of peak pressure during this velocity ramp. Additional information (radial distance at the start time, max P, max B, max n, min V, and max V) are also provided. The P and B maxima are defined as the peak value within the velocity increase, while the maximum of n is defined as the peak density within two days of the beginning of the velocity increase and two days following the velocity increase. All quantities are validated by eye to ensure the values picked are not outliers and that the peak in density corresponds to the density pileup region. The min V and max V are the minimum and maximum velocity in between the SIR and CIR start and end times identified. When data gaps are present during the velocity increase, the interface time and max P, B, and n are all set to "-", since these quantities cannot be reliably identified. An "ambiguous" flag is assigned using the criteria discussed in Section 2.4. Table \ref{tab1} indicates if each event has a corresponding observation at STA and/or L1, as well as if the structure was a CIR or not, that is to say if it had been observed to have persisted for a complete solar rotation. The determination of whether the structure is an SIR or CIR is done using corresponding observations. If the structure can be successfully tracked back in time by at least one solar rotation then it is considered a CIR, and, if not, it is deemed an SIR. Finally, the rightmost columns of Table \ref{tab1} list whether or not the structure was observed with a shock and if it was observed near a heliospheric current sheet (HCS) crossing

\subsection{Corresponding SIRs and CIRs}

\begin{figure*}
   \centering
   \includegraphics[width=18.25cm]{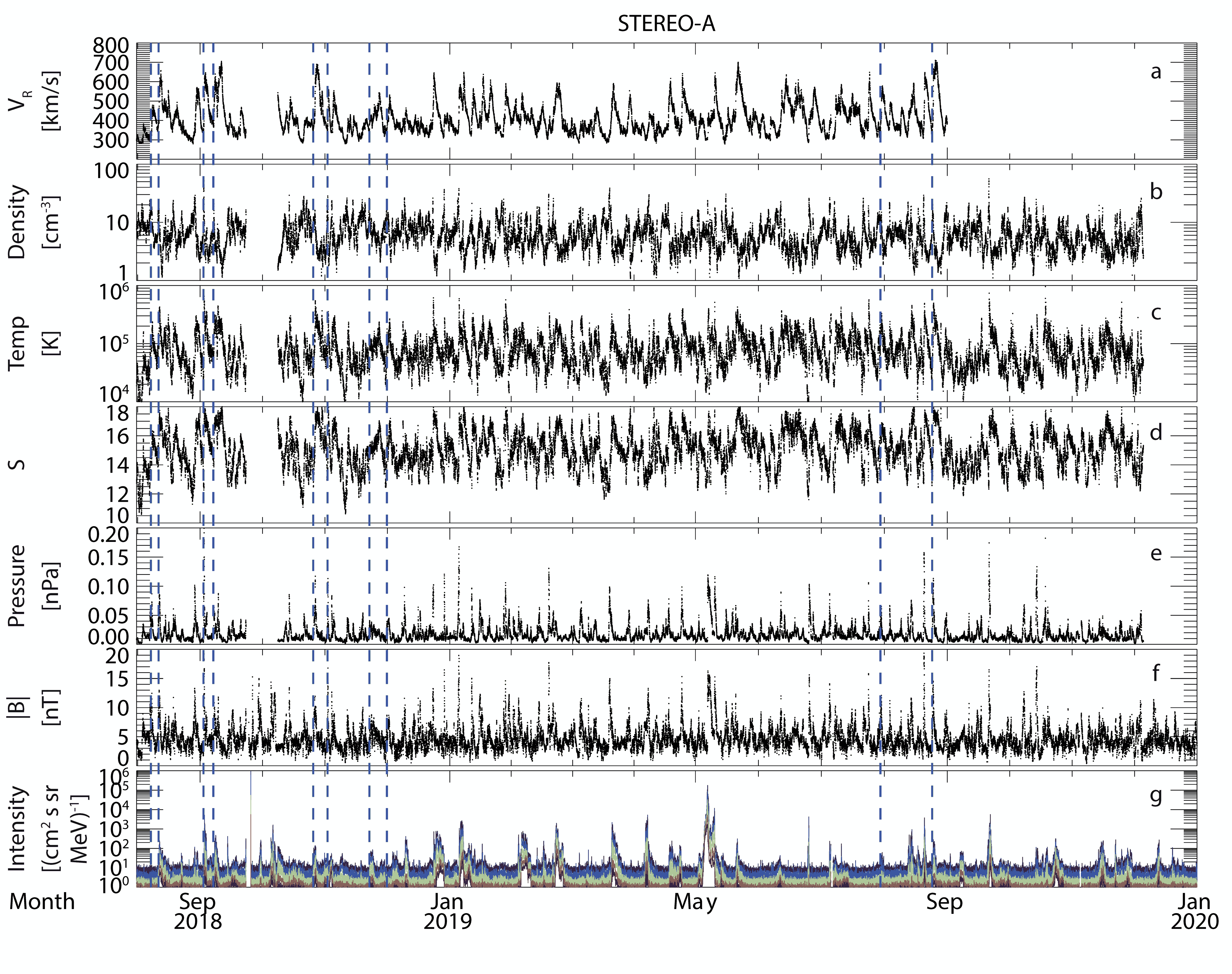}
      \caption{Overview of STA observations corresponding to SIRs and CIRs observed at PSP showing: (a) radial H\textsuperscript{+} velocity, (b) H\textsuperscript{+} density, (c) H\textsuperscript{+} temperature, (d) specific entropy argument, (e) sum of H\textsuperscript{+} plasma and magnetic pressures, (f) magnetic magnitude, and (g) suprathermal ion intensities for energies between 37.5 and 3000 keV. Vertical blue dashed lines illustrate the times of SIR and CIR observations that were also observed by PSP.
              }
         \label{fig2}
   \end{figure*}

To probe the evolution of SIRs and CIRs in both space (i.e., radial evolution) and time, each PSP SIR and CIR observation is tracked both backward and forward in time in a search for a corresponding observation at STA and L1 (ACE and Wind). Figure \ref{fig2} illustrates the STA observations over the same time period as the PSP observations in Figure \ref{fig1}, with vertical lines denoting the times of SIR and CIR observations corresponding to SIRs and CIRs identified at PSP. While the speed at which CIRs corotate has some variability (e.g., \citealp{Allen2020c}), the near-circular orbit of STA results in CIR observations to appear with a repeatable cadence approximately every 25 days. It is also clear that the elliptical orbit of PSP, along with its super-corotating speeds near perigee, can lead PSP to miss some SIR and CIR structures that are observed at STA. As such, the list of corresponding STA observations should not be taken as an exhaustive list of STA-observed 1 au SIRs and CIRs as has been compiled by others (e.g., \citealp{Jian2019}). Table \ref{tab2} lists the corresponding SIRs and CIRs identified in the STA data. As in Table \ref{tab1}, the first 11 columns mark the event times and characteristics. The following columns denote the PSP event number each STA observation corresponds to, whether or not there were also corresponding observations at L1, if the event at STA had been observed for at least one solar rotation prior to this observation, if the SIR and CIR had an associated shock, and if it was observed near a HCS crossing. If a CIR persists long enough, such as for STA events 1, 3, 5, and 7, then there can be multiple STA observations corresponding to multiple observations at PSP (PSP events 1 and 2 in this example).

\begin{figure*}
   \centering
   \includegraphics[width=18.25cm]{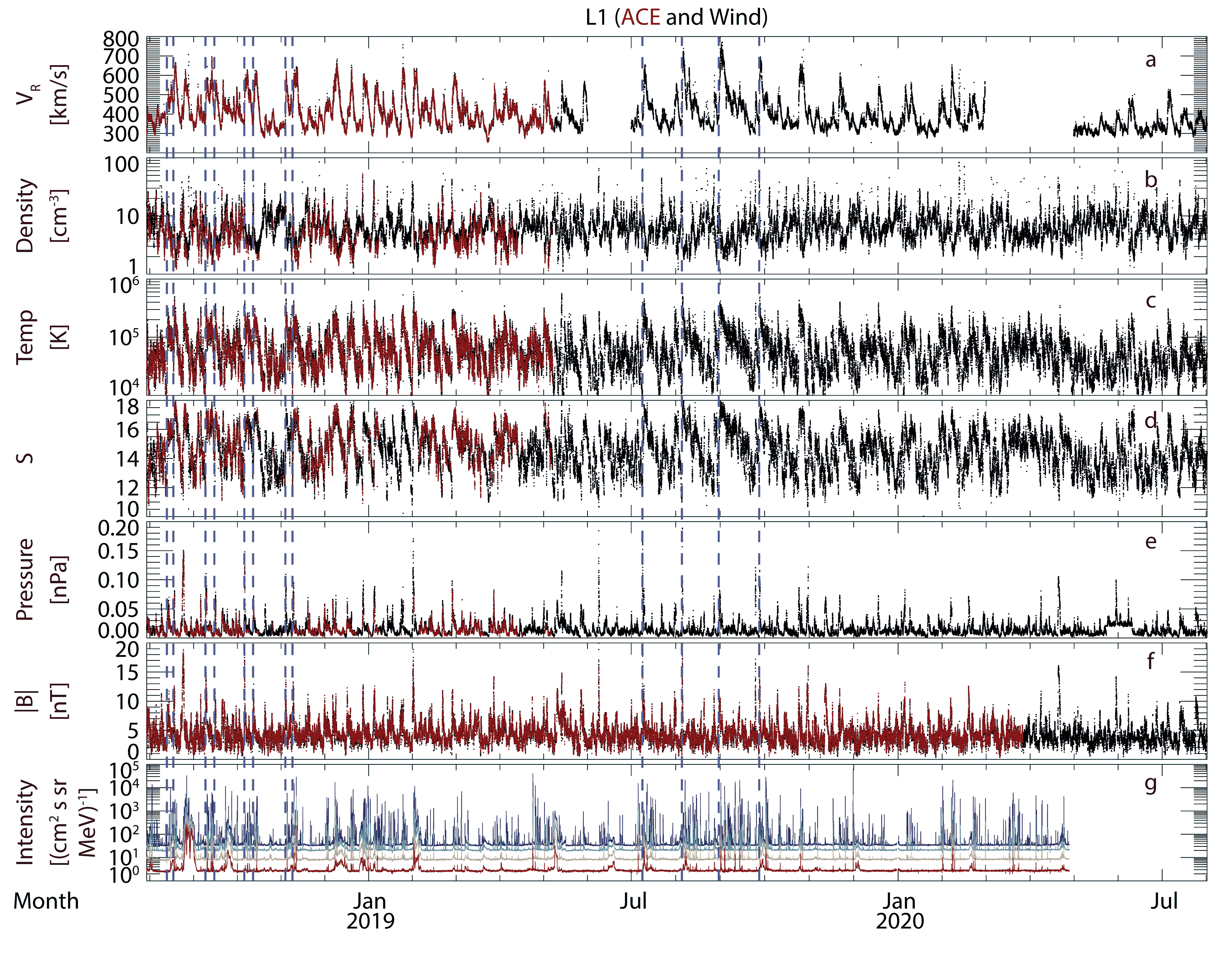}
      \caption{Overview of L1 observations (ACE in red and Wind in black for panels a-f) in the same format as Figure 2. Panel g illustrates the suprathermal ion intensities for the energy range of 56 to 3020 keV from ACE-EPAM. Vertical blue dashed lines illustrate the times of SIR and CIR observations that were also observed by PSP.
              }
         \label{fig3}
   \end{figure*}

Observations from ACE and Wind are shown in Figure \ref{fig3} in the same fashion as Figure \ref{fig2} (with ACE measurements in red and Wind measurements in black). As was the case for the STA events, the list of corresponding SIRs and CIRs identified in the ACE and Wind measurements should not be constructed as an exhaustive list at L1 as done by others (e.g., \citealp{Jian2006, Jian2011}). Table \ref{tab3} provides the catalog of corresponding SIRs and CIRs observations at L1 in the same fashion as Table \ref{tab2}. 

\subsection{Example of a clean event}

\begin{figure*}
   \centering
   \includegraphics[width=15cm]{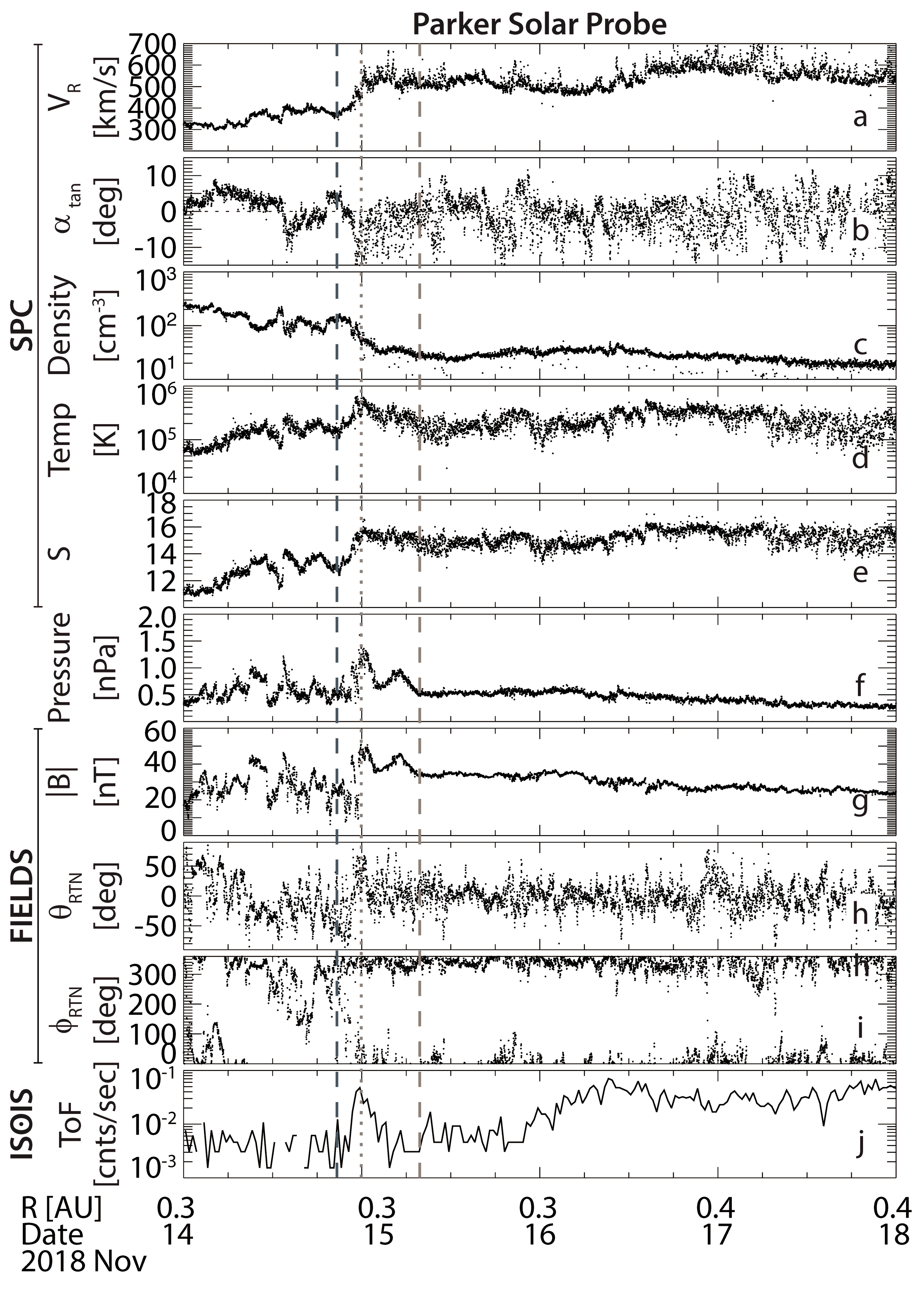}
      \caption{Example of a clean SIR and CIR event observed by PSP showing: (a) radial H\textsuperscript{+} velocity, (b) tangential H\textsuperscript{+} velocity deflection, (c) H\textsuperscript{+} density, (d) H\textsuperscript{+} temperature, (e) specific entropy argument, (f) sum of H\textsuperscript{+} plasma and magnetic pressure, (g) magnetic field magnitude, (h) magnetic $\Theta_{RTN}$, (i) magnetic $\Phi_{RTN}$, and (j) 30-586 keV suprathermal ion count rate. The blue and orange vertical dashed lines indicate where the radial velocity transitions from slow to fast solar wind. The red dotted line denotes the pressure peak within the velocity increase.
              }
         \label{fig4}
   \end{figure*}

An example of a "clean" SIR and CIR event observed by PSP is shown in Figure \ref{fig4}. This event exhibited all the classical CIR characteristics and is discussed in more detail in other studies \citep{Allen2020a, Joyce2020b, Schwadron2020}. The velocity increase is observed near 00:00 UT on 2018 November 15. The vertical blue and orange dashed lines denote the "start" and "stop" times listed in Table \ref{tab1}, respectively, while the vertical dotted red line indicates the interface (max P between the start and stop times). A clear tangential flow deflection occurred during the velocity ramp (Figure \ref{fig4}b). Additionally, the solar wind density is observed to have a pileup region upstream of the interface, followed by a lower density region downstream in the fast-solar wind (Figure \ref{fig4}c). The temperature and entropy are both higher in the high-speed stream as compared to the slow solar wind (Figure \ref{fig4}d and e). The pressure and magnetic field magnitude both peak at the interface (Figure \ref{fig4}f and g). For this event at PSP, the suprathermal ion intensity has an isolated peak at the SIR and CIR interface (Figure \ref{fig4}j), as well as a broader enhancement on 2018 November 16; these suprathermal ion enhancements are further discussed in \citet{Allen2020a}, and the ion spectra are investigated in detail by \citet{Joyce2020b}. Since all of the criteria for SIR and CIR selection are satisfied, and there are no data gaps for this event, this SIR and CIR is deemed a clean event (i.e., ambiguous flag of 0). Additionally, this structure was also observed by both STA and L1 (see Tables \ref{tab2} and \ref{tab3}, respectively), as well as by PSP at an earlier time (event 1 in Table \ref{tab1}). From the preceding observations at 1 au and at PSP, this event is identified as a CIR since the first observation of this structure was 98 days prior to the PSP observation on 2018 November 14-15 (i.e., more than one solar rotation). 

\begin{figure*}
   \centering
   \includegraphics[width=18.2cm]{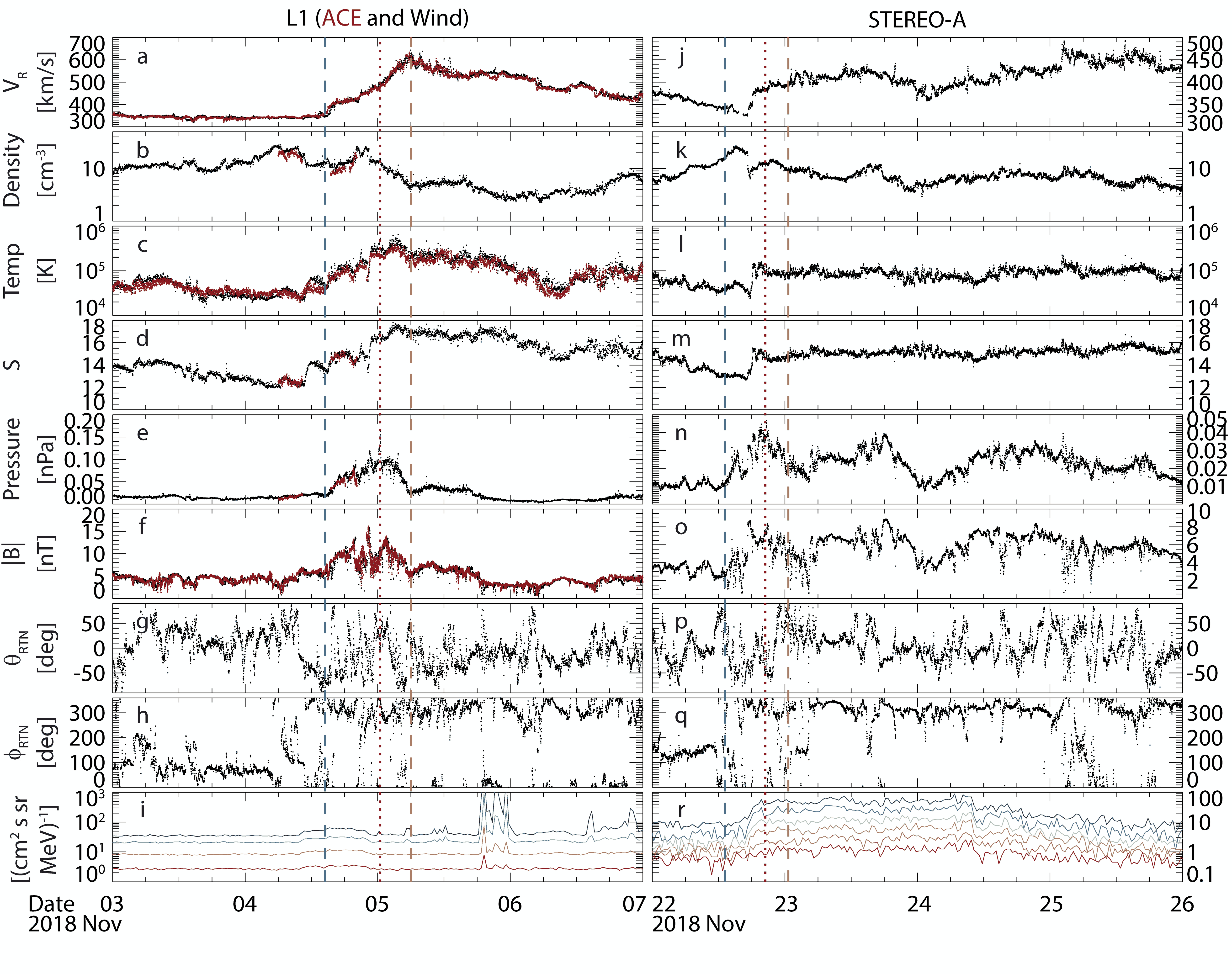}
      \caption{Nearest corresponding observations at L1 (left) and STEREO-A  (right) of the SIR and CIR observed by PSP in Figure 4. Panels show (a \& j) radial H\textsuperscript{+} velocity, (b \& k) H\textsuperscript{+} density, (c \& l) H\textsuperscript{+} temperature, (d \& m) specific entropy argument, (e \& n) sum of H\textsuperscript{+} plasma and magnetic pressure, (f \& o) magnetic field magnitude, (g \& p) magnetic $\Theta_{RTN}$, (h \& q) magnetic $\Phi_{RTN}$, and (i \& r) suprathermal ion intensities. Vertical lines indicate the start of the velocity increase (blue dashed), peak pressure during the velocity increase (red dotted), and the end of the velocity increase (orange dashed). 
              }
         \label{fig5}
   \end{figure*}

Figure \ref{fig5} illustrates the two temporally closest corresponding observations from L1 and STA. For this solar rotation, the CIR structure was first observed at L1 on 2018 November 4-5, and last observed at STA on 2018 November 22-23. Both L1 and STA detect a velocity enhancement, density pileup, temperature and entropy increase, and peak of P and B at the stream interface. However, notably, the suprathermal ion measurements at L1 do not show a large enhancement, while STA detects a clear enhancement following the interface and stretching into the high-speed stream. The differences between the three observations (L1, PSP, and STA) illustrate the importance of multipoint measurements to better understand the evolution of large-scale structures in the heliosphere. As discussed in \citet{Allen2020a}, the differences in the suprathermal ion populations at and near the stream interface may be due to the relative contribution of locally-accelerated populations versus the inward transport of shock-accelerated ions (see also \citealp{Schwadron1996, Filwett2019}). Through investigating the PSP SIR and CIR structures in the context of 1 au measurements, insight into both the radial and temporal evolution can be gathered. 

\subsection{Example of an ambiguous event}

\begin{figure*}
   \centering
   \includegraphics[width=15cm]{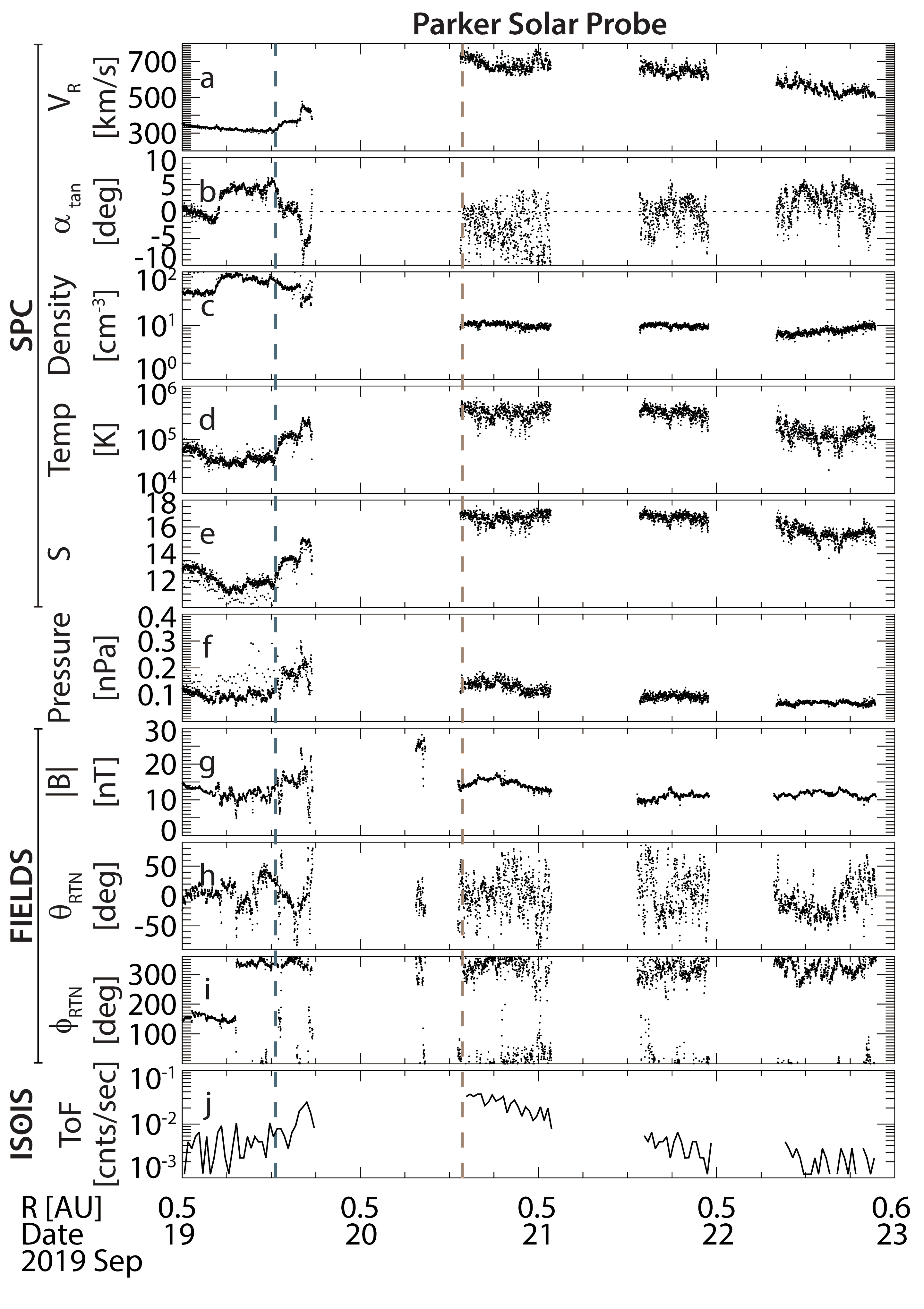}
      \caption{Example of an ambiguous SIR and CIR event observed by PSP in the same format as Figure 4.
              }
         \label{fig6}
   \end{figure*}

An example of an ambiguous SIR and CIR event observed by PSP is shown in Figure \ref{fig6}. This event has previously been studied in depth by \citet{Allen2020b}. Due to data gaps caused by Ka-band downlinking from the spacecraft, only fragments of this interval were collected. From the intervals with data, however, the beginning of the velocity increase is observed along with a tangential flow deflection, density pileup, and enhancement in temperature and entropy from slow to fast wind. As most of the velocity increase is missing, it is not possible to identify the interface, which would have occurred sometime between 18:00 UT on 2019 September 19 to 12:00 UT on 2019 September 20 (Figure \ref{fig6}a). However, the suprathermal ions did exhibit an enhancement during the start of the high-speed stream (Figure \ref{fig6}j). Given the available data, this event is categorized as an SIR and CIR, and so it is included in our catalog of events (Table \ref{tab1}). However, due to data gaps occurring during the velocity enhancement, this event is flagged as ambiguous (ambiguous counter of 2). It should also be noted that this structure persisted for multiple solar rotations as observed at 1 au (Tables \ref{tab2} and \ref{tab3}) and as such is identified as a CIR. While these observations from PSP are not able to provide a full picture of the CIR, such as information of the peak pressure at the interface, these types of observations can be nonetheless scientifically valuable. As an example, for the event shown in Figure \ref{fig6} (Event 10 in Table \ref{tab1}), the CIR was observed when PSP and STA were nearly radially aligned, enabling an examination of the radial variation in the CIR properties that were measured by PSP. 

\begin{figure*}
   \centering
   \includegraphics[width=18.2cm]{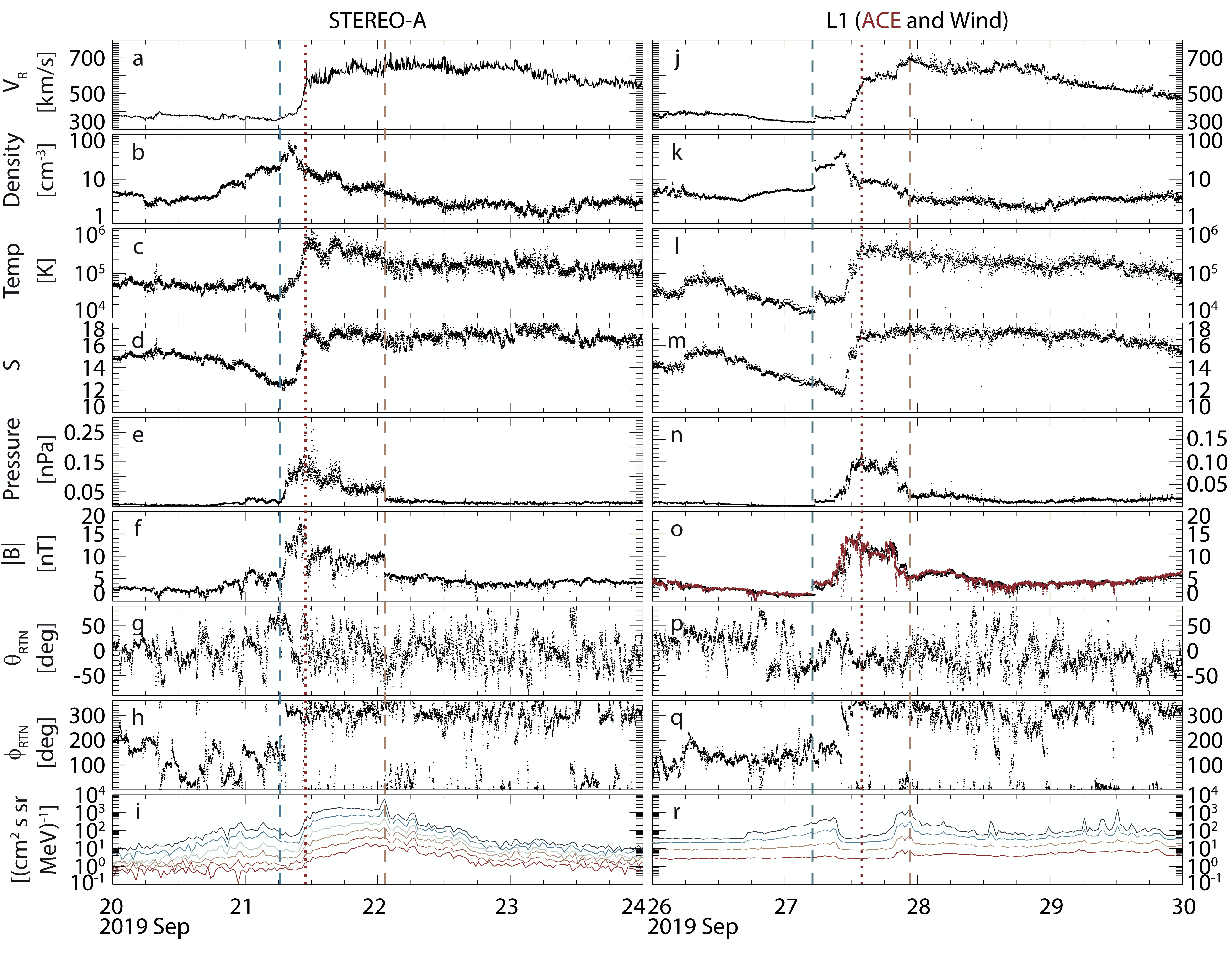}
      \caption{Closest corresponding observations of the SIR and CIR observed by PSP in Figure 6 at STEREO-A (left) and L1 (right) shown in the same format as Figure 5.
              }
         \label{fig7}
   \end{figure*}

Figure \ref{fig7} shows the most recent L1 and STA observations corresponding to the PSP event shown in Figure \ref{fig6} (all observations of this CIR are discussed in \citealp{Allen2020b}). The CIR was observed at STA approximately 1.77 days after the PSP observations and at L1 \textasciitilde7.5 days after the PSP observation. All of the signatures of the CIR are seen to be very similar between the three observations, indicating that, in contrast to the event in Figure \ref{fig4} and Figure \ref{fig5}, the structure observed in late September 2019 is not evolving much between observations by the different spacecraft. This static nature of the mature CIR shown in Figures \ref{fig6} and \ref{fig7} allows for investigations of the radial variations of CIRs and their associated energetic particles (see \citealp{Allen2020b}).

\subsection{Initial statistical trends}

To investigate the evolution of SIR and CIR properties as a function of heliocentric distance, the maximum density, peak pressure, and peak magnetic field magnitude for the events are scaled by radial distance to 1 au. Following \citet{Perrone2019}, the density is scaled as $n \propto R^{-1.96}$ and magnetic field magnitude as $B \propto R^{-1.59}$. The scaled pressure was computed by first scaling the temperature using $T \propto R^{-0.93}$ \citep{Perrone2019} as well as the scaled density and magnetic field magnitude to recompute the pressure using equation 3. Due to data gaps in ambiguous events potentially leading to a misidentification of the max density and peak magnetic field magnitude and pressure, only clean PSP events are used for this initial statistical study. Scaling the observations by radial distance allows for an investigation of radial variations due to the interaction of the fast and slow speed streams (i.e., this scaling removes the underlying systematic radial variations due to processes such as volumetric expansion that would otherwise dominate the radial trends). 

\begin{figure}
   \centering
   \includegraphics[width=9cm]{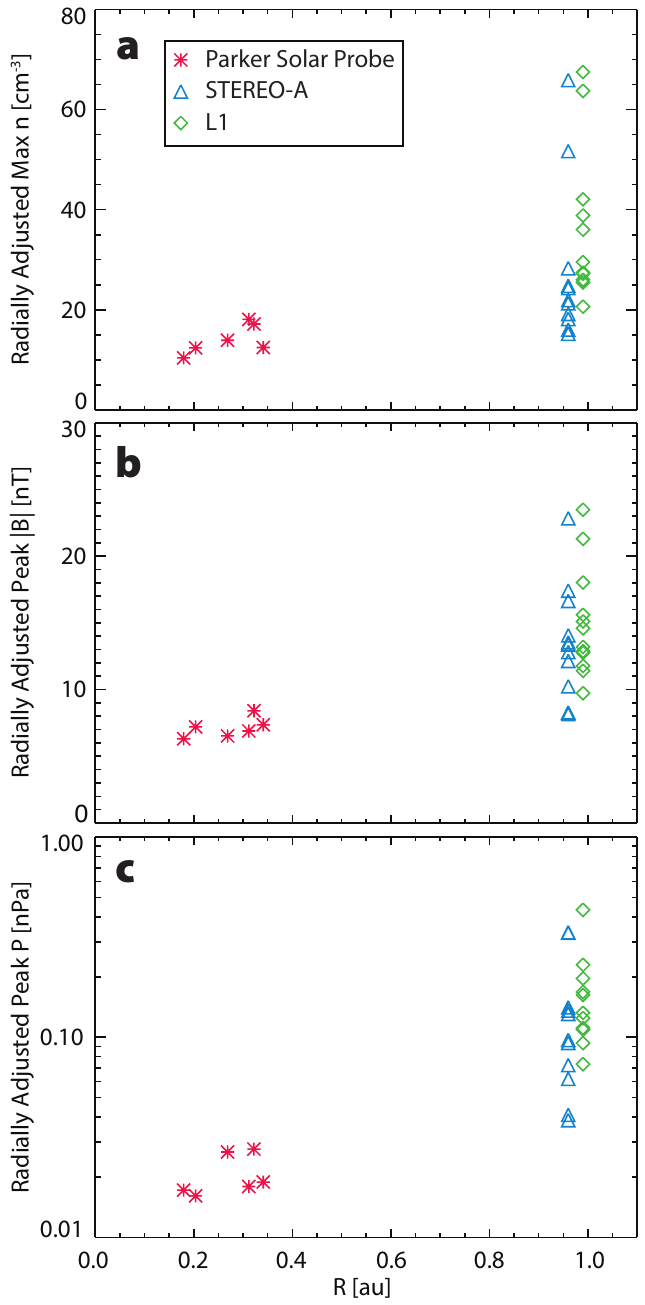}
      \caption{Radial profile of (a) adjusted max density, (b) adjusted peak magnetic field magnitude, and (c) adjusted peak pressure. Values were adjusted to remove bulk radial variations following the radial relations in \citet{Perrone2019}.
              }
         \label{fig8}
   \end{figure}

Figure \ref{fig8} shows the radial evolution of the scaled max density, magnetic field magnitude, and pressure. While there is considerable variation in the SIR and CIR properties at 1 au, they are often higher than the six clean PSP events. These results are consistent with an earlier interpretation of Helios observations (e.g., \citealp{Richter1986}), where the SIR and CIR structures develop the density pile-up and compression of the magnetic field, resulting in a pressure enhancement that grows with radial distance. However, because of the limited number of clean events in the inner heliosphere, as well as the spread in values at 1 au and known temporal variations in the SIRs and CIRs, additional observations are required before statistically meaningful relations can be extrapolated from these observations. These early statistical trends do indicate that future investigations using PSP, STEREO-A, L1 observatories, and eventually Solar Orbiter will allow for further investigations in the radial evolution of SIRs and CIRs.

\section{Summary}

 SIRs and CIRs are large-scale, and often long-lived structures that are an important aspect of the solar wind and dynamics in interplanetary space. Understanding how these structures evolve in the inner heliosphere as they form and persist for multiple solar rotations requires a robust database of observations at small heliospheric distances and complementary measurements from other observatories at different locations and successive points in time. Such a living catalog of SIR and CIR events observed by PSP with corresponding observations at STA and L1 (ACE and Wind), and the criteria for inclusion in the catalog, are presented here. Future iterations of the catalog will also include the identification of corresponding events at Solar Orbiter. These catalogs of SIRs and CIRs will be updated and published on the PSP Science Gateway (https://sppgway.jhuapl.edu/event\_list/) concurrent to PSP public data releases. 
 
 While adhering to stringent identification criteria, the catalog of PSP events includes intervals where measurements are not fully available for complete characterization of the stream interface and parts of the upstream or downstream regions. However, as demonstrated by the two different PSP events highlighted here, certain features of the SIR and CIR structures can still be analyzed in depth. As also illustrated by the two examples of SIR and CIR events shown here, corresponding observations, even for incomplete measurement intervals, allow for studies of both temporal (longitudinal) and radial evolution of SIRs and CIRs between observatories in a way that has not been possible since the Helios era. This is further demonstrated by the initial analysis of the radial variation of certain SIR and CIR properties also included here, which shows evidence of systematic increases in bulk characteristics at the structure interface with increasing heliocentric distance. Additionally, the advanced instrumentation of modern missions, along with latitudinal separation provided by Solar Orbiter, will allow for more detailed investigations of SIR and CIR structures and associated energetic plasma populations.
 
\begin{acknowledgements}
This work was supported by NASA's Parker Solar Probe Mission, contract NNN06AA01C and by NASA's ACE grant NNX17AC05G. C.M. thanks the Austrian Science Fund (FWF): P31659-N27. The ISOIS data and visualization tools are available to the community at: https://spacephysics.princeton.edu/missions-instruments/isois ; data are also available via the NASA Space Physics Data Facility (https://spdf.gsfc.nasa.gov/) and from https://sppgway.jhuapl.edu/event\_list. The STEREO SEPT data are available at http://www2.physik.unikiel.de/stereo/data/sept/, and STEREO magnetic field and plasma data can be found at the STEREO Science Center at https://stereo-ssc.nascom.nasa.gov. Wind data are available at https://cdaweb.gsfc.nasa.gov/. The ACE data are publicly available at the ACE mission website at http://www.srl.caltech.edu/ACE/. WSA-ENLIL movies of PSP orbits are available at http://helioweather.net/missions/psp/. Living versions of the catalog presented in this paper are available and kept up to date on both the Parker Solar Probe Science Gateway (https://sppgway.jhuapl.edu/event\_list/). 
\end{acknowledgements}


\begin{sidewaystable*}
\caption{PSP SIR and CIR events.}\label{tab1}
\centering
\begin{tabular}{cccccccccccccccc} 
\hline\hline             
\# & Start Time & End Time 
& Time of & $R$ & Max $P$ & Max $B$ & Max $n_p$ & Min $V$ & Max $V$ & Ambiguous & STA & L1 & CIR & Shock & HCS \\
 & & & max $P$ & [au] & [nPa] & [nT] & [cm\textsuperscript{-3}] & [km/s] & [km/s] & Marker & & & & &\\
\hline
  1 & 2018-10-03 & 2018-10-06 & -- & 0.72 & -- & -- & -- & 342 & 574 & 2 & Y & Y & Y & -- & Y\\
  & 22:00:00 & 08:00:00 & &&&&&&&&& & &&\\
  2 & 2018-11-14 & 2018-11-15 & 2018-11-14 & 0.32 & 1.368 & 50 & 165 & 346 & 666 & 0 & Y & Y & Y & N & Y\\
  & 20:40:00 & 07:25:00 & 23:56:30 &&&&&&&&&& & &\\
3 & 2018-11-26 & 2018-11-27 & -- & 0.54 & -- & -- & -- & 360 & 662 & 2 & Y & N & N & -- & N\\
  & 15:05:00 & 06:10:00 & &&&&&&&&& & &&\\
4 & 2019-04-06 & 2019-04-07 & 2019-04-06 & 0.17 & 3.901 & 97 & 322 & 211 & 498 & 0 & N & N & N & N & N\\
  & 20:20:00 & 01:25:00 & 21:55:30 &&&&&&&&&& & &\\
5 & 2019-04-16 & 2019-04-17 & -- & 0.38 & -- & -- & -- & 320 & 717 & 2 & N & N & N & -- & --\\
  & 19:40:00 & 18:00:00 & &&&&&&&&& & &&\\
6 & 2019-08-22 & 2019-08-23 & 2019-08-23 & 0.34 & 0.713 & 41 & 107 & 283 & 457 & 0 & N & N & N & N & N\\
  & 22:00:00 & 04:40:00 & 00:59:30 &&&&&&&&&& & &\\
7 & 2019-08-24 & 2019-08-24 & 2019-08-24 & 0.31 & 0.923 & 44 & 186 & 306 & 431 & 0 & N & N & N & N & N\\
  & 06:00:00 & 15:30:00 & 08:37:30 &&&&&&&&&& & &\\
8 & 2019-08-26 & 2019-08-26 & 2019-08-26 & 0.27 & 1.516 & 53 & 193 & 315 & 476 & 0 & N & N & N & N & N\\
  & 04:55:00 & 11:50:00 & 06:22:30 &&&&&&&&& & &&\\
9 & 2019-08-29 & 2019-08-29 & 2019-08-29 & 0.20 & 3.900 & 90 & 298 & 267 & 476 & 0 & N & N & N & N & N\\
  & 05:30:00 & 13:50:00 & 07:03:30 &&&&&&&&&& & &\\
10 & 2019-09-19 & 2019-09-20 & -- & 0.50 & -- & -- & -- & 315 & 751 & 2 & Y & Y & Y & -- & Y\\
  & 12:35:00 & 13:45:00 & &&&&&&&&& & &&\\
11 & 2020-05-22 & 2020-05-22 & -- & 0.48 & -- & -- & -- & 363 & 533 & 2 & N & N & N & N & Y\\
  & 05:05:00 & 08:05:00 & &&&&&&&&& & &\\
\end{tabular}
\tablefoot{Event 1 and 2 are the same CIR structure.}
\end{sidewaystable*}

\begin{sidewaystable*}
\caption{Corresponding STA SIR and CIR events.}\label{tab2}
\centering
\begin{tabular}{cccccccccccccccc} 
\hline\hline             
\# & Start Time & End Time 
& Time of & $R$ & Max $P$ & Max $B$ & Max $n_p$ & Min $V$ & Max $V$ & Ambiguous & PSP & L1 & CIR & Shock & HCS\\
 & & & max $P$ & [au] & [nPa] & [nT] & [cm\textsuperscript{-3}] & [km/s] & [km/s] & Marker & Event \# & & \\
\hline
1 & 2018-08-08 & 2018-08-08 & 2018-08-08 & 0.96 & 0.103 & 14 & 19.7 & 305 & 479 & 0 & 1; 2 & Y & N & N & Y\\
  & 00:00:00 & 18:00:00 & 11:23:30 &&&&&&&&&&&&\\
2 & 2018-08-11 & 2018-08-12 & 2018-08-12 & 0.96 & 0.106 & 14 & 26.4 & 364 & 683 & 0 & 3 & Y & N & N & Y\\
  & 14:40:00 & 23:40:00 & 06:21:30 &&&&&&&&&&&&\\
3 & 2018-09-02 & 2018-09-03 & 2018-09-02 & 0.96 & 0.352 & 24 & 56.1 & 336 & 651 & 0 & 1; 2 & Y & Y & N & Y\\
  & 16:00:00 & 06:00:00 & 22:54:30 &&&&&&&&&&&&\\
4 & 2018-09-07 & 2018-09-08 & 2018-09-07 & 0.96 & 0.070 & 11 & 23.1 & 357 & 631 & 0 & 3 & Y & Y & N & Y\\
  & 12:00:00 & 05:40:00 & 20:00:30 &&&&&&&&&&&&\\
5 & 2018-10-26 & 2018-10-27 & 2018-10-27 & 0.96 & 0.144 & 13 & 17.4 & 347 & 700 & 0 & 1; 2 & Y & Y & N & Y\\
  & 06:30:00 & 20:50:00 & 07:28:30 &&&&&&&&&&&&\\
6 & 2018-11-02 & 2018-11-02 & 2018-11-02 & 0.96 & 0.152 & 18 & 30.7 & 345 & 556 & 0 & 3 & Y & Y & N & Y\\
  & 06:00:00 & 18:10:00 & 11:21:30 &&&&&&&&&&&&\\
7 & 2018-11-22 & 2018-11-23 & 2018-11-22 & 0.96 & 0.047 & 9 & 26.9 & 324 & 406 & 0 & 1; 2 & Y & Y & N & Y\\
  & 13:35:00 & 00:35:00 & 20:40:30 &&&&&&&&&&&&\\
8 & 2018-12-01 & 2018-12-01 & 2018-12-01 & 0.96 & 0.042 & 9 & 20.8 & 321 & 430 & 0 & 3 & Y & Y & N & Y\\
  & 05:55:00 & 09:55:00 & 06:13:30 &&&&&&&&&&&&\\
9 & 2019-07-30 & 2019-07-31 & 2019-07-30 & 0.96 & 0.078 & 14 & 16.5 & 323 & 584 & 0 & 10 & Y & N & N & Y\\
  & 07:50:00 & 05:50:00 & 21:02:30 &&&&&&&&&&&&\\
10 & 2019-08-24 & 2019-08-25 & 2019-08-25 & 0.96 & 0.151 & 15 & 23.8 & 324 & 697 & 0 & 10 & Y & Y & N & Y\\
  & 13:00:00 & 18:25:00 & 05:05:30 &&&&&&&&&&&&\\
11 & 2019-09-21 & 2019-09-22 & 2019-09-21 & 0.96 & 0.370 & 19 & 71.5 & 349 & 695 & 0 & 10 & Y & Y & N & Y\\
  & 06:20:00 & 01:05:00 & 10:55:30 &&&&&&&&&&&&\\
\end{tabular}
\end{sidewaystable*}

\begin{sidewaystable*}
\caption{Corresponding L1 SIR and CIR events.}\label{tab3}
\centering
\begin{tabular}{cccccccccccccccc} 
\hline\hline             
\# & Start Time & End Time 
& Time of & $R$ & Max $P$ & Max $B$ & Max $n_p$ & Min $V$ & Max $V$ & Ambiguous & PSP & STA & CIR & Shock & HCS\\
 & & & max $P$ & [au] & [nPa] & [nT] & [cm\textsuperscript{-3}] & [km/s] & [km/s] & Marker & Event \# & & \\
\hline
1 & 2018-08-14 & 2018-08-16 & 2018-08-15 & 0.99 & 0.073 & 10 & 28.0 & 308 & 550 & 0 & 1; 2 & Y & N & N & Y\\
  & 22:00:00 & 00:10:00 & 17:33:30 &&&&&&&&&&&&\\
2 & 2018-08-19 & 2018-08-20 & 2018-08-20 & 0.99 & 0.111 & 13 & 26.0 & 417 & 674 & 0 & 3 & Y & N & N & Y\\
  & 16:20:00 & 08:40:00 & 05:50:30 &&&&&&&&&&&&\\
3 & 2018-09-10 & 2018-09-11 & 2018-09-10 & 0.99 & 0.125 & 15 & 27.8 & 359 & 658 & 0 & 1; 2 & Y & Y & N & Y\\
  & 09:50:00 & 12:20:00 & 21:39:30 &&&&&&&&&&&&\\
4 & 2018-09-16 & 2018-09-17 & 2018-09-17 & 0.99 & 0.230 & 12 & 36.8 & 380 & 593 & 0 & 3 & Y & Y & N & Y\\
  & 23:30:00 & 18:10:00 & 13:25:30 &&&&&&&&&&&&\\
5 & 2018-10-07 & 2018-10-07 & 2018-10-07 & 0.99 & 0.169 & 18 & 65.0 & 354 & 597 & 0 & 1; 2 & Y & Y & N & Y\\
  & 07:30:00 & 21:20:00 & 12:52:30 &&&&&&&&&&&&\\
6 & 2018-10-13 & 2018-10-14 & 2018-10-13 & 0.99 & 0.094 & 13 & 39.6 & 323 & 614 & 0 & 3 & Y & Y & N & Y\\
  & 08:00:00 & 16:20:00 & 16:50:30 &&&&&&&&&&&&\\
7 & 2018-11-04 & 2018-11-05 & 2018-11-05 & 0.99 & 0.163 & 16 & 26.5 & 345 & 643 & 0 & 1; 2 & Y & Y & N & Y\\
  & 14:30:00 & 06:00:00 & 00:27:30 &&&&&&&&&&&&\\
8 & 2018-11-09 & 2018-11-10 & 2018-11-10 & 0.99 & 0.125 & 13 & 21.0 & 380 & 614 & 0 & 3 & Y & Y & N & Y\\
  & 12:00:00 & 15:40:00 & 08:37:30 &&&&&&&&&&&&\\
9 & 2019-07-08 & 2019-07-10 & 2019-07-08 & 0.99 & 0.197 & 22 & 68.9 & 298 & 653 & 0 & 10 & Y & N & Y & Y\\
  & 17:00:00 & 07:50:00 & 22:13:30 &&&&&&&&&&&&\\
10 & 2019-08-04 & 2019-08-05 & 2019-08-05 & 0.99 & 0.433 & 24 & 30.1 & 320 & 733 & 0 & 10 & Y & Y & N & Y\\
  & 23:00:00 & 21:00:00 & 08:01:30 &&&&&&&&&&&&\\
11 & 2019-08-30 & 2019-08-31 & 2019-08-31 & 0.99 & 0.109 & 12 & 27.9 & 373 & 736 & 0 & 10 & Y & Y & N & Y\\
  & 08:25:00 & 13:10:00 & 01:06:30 &&&&&&&&&&&&\\
12 & 2019-09-27 & 2019-09-27 & 2019-09-27 & 0.99 & 0.132 & 15 & 42.9 & 340 & 706 & 0 & 10 & Y & Y & Y & Y\\
  & 05:00:00 & 22:40:00 & 13:54:30 &&&&&&&&&&&&\\

\end{tabular}
\end{sidewaystable*}

\end{document}